\documentclass[twocolumn]{openjournal}

\usepackage{amsmath}
\usepackage{bm}
\usepackage{amsfonts}
\usepackage{graphicx}
\usepackage{hyperref}
\usepackage{color}

\hypersetup{
colorlinks=true,
linkcolor=blue,
linktoc=page,
citecolor=blue,
urlcolor=blue}

\begin{document}

\title{\bf Gravitational Wave Direct Detection Does Not Constrain the Tensor Spectral Index at CMB Scales}

\newcommand{\FIRSTAFF}{\affiliation{Department of Physics, University at Buffalo, Buffalo, NY 14260, USA}}

\author{William H. Kinney}
\email[Electronic address: ]{whkinney@buffalo.edu}
\FIRSTAFF
\date{\today}
\begin{abstract}
I discuss constraints on the power spectrum of primordial tensor perturbations from a combination of Cosmic Microwave Background (CMB) measurements and the gravitational wave direct detection experiments LIGO/Virgo and DECIGO. There are two main points: (1) Inflation predicts an approximately power-law form for the primordial tensor spectrum, but makes no prediction for its amplitude. Given that neither Planck nor LIGO/Virgo has actually detected primordial tensor modes, it is trivially true that no model-independent constraint on the \textit{slope} of the tensor power spectrum is possible with current data. (2) CMB and LIGO/Virgo scales differ by more than 19 orders of magnitude, and 16 for DECIGO. I show that a power-law extrapolation from CMB to direct detection frequencies overestimates the amplitude of primordial tensor modes by as much as two orders of magnitude relative to an ensemble of realistic single-field inflation models. Moreover, the primordial tensor amplitude at direct detection scales is mostly uncorrelated with the tensor spectral index at CMB scales, and any constraint is strongly dependent on the specific form of the inflationary potential. 
\end{abstract}

\maketitle

\section{Introduction}

Cosmological inflation \cite{Starobinsky:1980te,Sato:1981ds,Sato:1980yn,Kazanas:1980tx,Guth:1980zm,Linde:1981mu,Albrecht:1982wi} predicts a nearly scale-invariant power-law spectrum of primordial gravitational waves \cite{Starobinsky:1979ty,Mukhanov:1981xt,Mukhanov:2003xw,Linde:1983gd,Hawking:1982cz,Hawking:1982my,Starobinsky:1982ee,Guth:1982ec,Bardeen:1983qw}, extending from superhorizon scales to wavelengths potentially accessible to future direct detection experiments \cite{Turner:1996ck,Ungarelli:2005qb,Cooray:2005xr,Smith:2005mm,Smith:2006xf,Chongchitnan:2006pe,Friedman:2006zt,Kawamura:2011zz,Kawamura:2020pcg,Caligiuri:2014ola,Caprini:2018mtu}. 

The Planck measurement of the Cosmic Microwave Background constrains the amplitude primordial gravitational wave (or \textit{tensor}) perturbations on scales comparable to the current Hubble length, 
\begin{equation}
    a_0 H_0 = 2.248 \times 10^{-4}\ \mathrm{Mpc}^{-1},
\end{equation}
placing an upper bound on the ratio of primordial gravitational wave perturbations $P_T$ to curvature perturbations $P_{\mathcal{R}}$ (the \textit{tensor/scalar ratio}), measured at a pivot scale of $k = 0.002\ \mathrm{Mpc}^{-1}$ of 
\begin{equation}
    r_{0.002} \equiv \frac{P_T}{P_{\mathcal{R}}}\bigg\vert_{k = 0.002\ \mathrm{Mpc}^{-1}} < 0.06.
\end{equation}
This bound is from Planck in combination with the BICEP/Keck measurement of CMB polarization and Baryon Acoustic Oscillation (BAO) measurements \cite{Aghanim:2018eyx}. (I adopt the Planck best-fit value for the Hubble constant of $H_0 = 67.4\ \mathrm{km/s/Mpc}$ throughout.) Inflation predicts the production of primordial gravitational waves with amplitude \begin{equation}
    r  = 16 \epsilon,
\end{equation}
where $\epsilon$ is the first slow roll parameter, defined in terms of the Hubble parameter as
\begin{equation}
    \epsilon \equiv \frac{1}{H}\frac{d H}{d N}.
\end{equation}
$d N \equiv - H dt$ is the number of e-folds of expansion, $a\left(N\right) \propto e^{-N}$. The spectrum of primordial gravity waves is a power-law, 
\begin{equation}
\label{eq:PT}
    P_T = \frac{2 H^2}{\pi^2 M_{\mathrm{P}}^2}\bigg\vert_{k = a H} \propto k^{n_T},
\end{equation}
where $k = a H$ indicates evaluation when a wavenumber $k$ exits the horizon during inflation. In the case of single-field inflation, the tensor spectral spectral index is also determined by $\epsilon$, 
\begin{equation}
\label{eq:consistency}
    n_T = - r / 8 = -2 \epsilon.
\end{equation}
Because this \textit{consistency relation} is a prediction of single-field inflation, constraining the tensor spectral index $n_T$ is of great interest. A constraint on the slope of the tensor spectrum requires measurement of the tensor amplitude at multiple scales: direct detection experiments such as LIGO/Virgo are an ideal candidate for such a measurement, and a number of papers in the recent literature have derived bounds on $n_T$ from Planck in combination with bounds on the contribution of gravity waves to the number of relativistic degress of freedom $N_{eff}$ and from LIGO/Virgo \cite{Lasky:2015lej,Cabass:2015jwe,Akrami:2018odb,Giare:2019snj,Tanin:2020qjw}. Bounds from $N_{eff}$ depend on the \textit{integrated} density in gravitational waves over all scales, but bounds from LIGO/Virgo depend on extrapolation to high frequency. In this paper, I focus on the latter case.

A few general points can be made at the outset: First is that \textit{neither} Planck nor LIGO/Virgo has actually detected primordial gravitational waves. Both measurements produce only upper bounds on the primordial tensor amplitude, and the signal from slow-roll inflation is expected to be many orders of magnitude below LIGO/Virgo sensitivity \cite{Aasi:2014zwg}. It therefore follows trivially that no model-independent constraint on $n_T$ is possible: if you do not know the value of a function at any point, quoting a constraint on its slope is obviously nonsensical. A model-dependent constraint can be derived, \textit{e.g.} by assuming a value of $r$ at CMB scales, and then placing a constraint on $n_T$ from bounds on $N_{eff}$ or from the non-detection of primordial tensor modes by LIGO/Virgo. This latter option, however, requires a substantial extrapolation. LIGO/Virgo is sensitive to gravitational waves between roughly 10 and 1000 Hz, with maximum sensitivity at a frequency of about 100 Hz. Translated into wavenumber, this frequency corresponds to $k = 6.47 \times 10^{16}\ \mathrm{Mpc}^{-1}$. Comparing the tensor amplitude  from a CMB pivot scale of $k = 0.002\ \mathrm{Mpc}^{-1}$ to LIGO/Virgo scales therefore requires an extrapolation of more than 19 orders of magnitude! It is not at all obvious that the assumption of a power-law for the tensor spectrum -- which itself presupposes an inflationary origin for the perturbations -- is even approximately valid over such a large range of scales. 

Primordial tensor modes are created during inflation by the redshift of quantum fluctuations to superhorizon scales, with an amplitude which is determined by the expansion rate at the time a particular wavenumber crosses the horizon scale (Eq. \ref{eq:PT}). It is convenient to express the condition for horizon crossing in terms of the number of e-folds before the end of inflation at which a mode exits the horizon. I adopt the usual convention for the number of e-folds $N$, such that $N = 0$ denotes the end of inflation and the onset of reheating, which $N$ increasing going back in time, and earlier into the inflationary epoch. Long wavelength modes exit the horizon earlier in inflation, and short wavelength modes exit the horizon later. Because of this, we expect that wavelengths corresponding to direct detection scales will exit the horizon near the end of inflation.  For a wavenumber $k$, the number of e-folds at which it crosses outside the horizon during inflation is
\begin{eqnarray}
N_k = &&- \ln\left(\frac{k}{a_0 H_0}\right) + \ln\left(\frac{H_*}{H_e}\right) + \left(\frac{1 + 3 {\bar w}}{2}\right) \left\vert N_{RH} \right\vert\cr 
&& + \ln\left(\frac{T_{RH}}{T_{eq}}\right) + \frac{1}{3} \ln\left(\frac{g_{*S}(T_{RH})}{g_{*S}(T_{eq})}\right)\cr 
&&+ \ln\left(\frac{(a H)_{eq}}{(a H)_0}\right),
\end{eqnarray}
where $a_0 H_0$ is the inverse of the comoving horizon size in the current universe, and subscript $eq$ indicates quantities evaluate at matter/radiation equality. Here $N_{RH}$ is the number of e-folds of expansion during reheating, and $\bar w$ is the average equation of state during the reheating period, before the universe becomes radiation-dominated.  $H_*$ is the value of the Hubble parameter during inflation when the mode exited the horizon, and $H_e$ is the Hubble parameter at the end of inflation. The factor $g_{*S}$ is the entropy in relativistic degrees of freedom. Taking Planck best-fit values for the matter and Dark Energy density gives
\begin{equation}
    \ln\left(\frac{(a H)_{eq}}{(a H)_0}\right) = 3.839.
\end{equation}
I assume matter-domination during the reheat period, ${\bar w} = 0$, so that 
\begin{equation}
\left(\frac{1 + 3 {\bar w}}{2}\right) \left\vert N_{RH} \right\vert = - \frac{2}{3} \ln\left(\frac{T_{RH}}{\rho_e^{1/4}}\right),
\end{equation}
where $\rho_e$ is the energy density at the end of inflation. For the purposes of the present discussion, it is sufficient to simplify by taking $H_* \simeq H_e$, and $g_{*S}\left(T_{RH}\right) \simeq 100$, so that 
\begin{eqnarray}
    \label{eq:NvsT}
    N\left(k\right) = &&- \ln\left(\frac{k}{a_0 H_0}\right) + \frac{1}{3} \ln\left(\frac{T_{RH}}{10^{15}\ \mathrm{GeV}}\right)\cr  &&
    + \frac{2}{3} \ln\left(\frac{\Lambda}{10^{15}\ \mathrm{GeV}}\right) + 60.4,
\end{eqnarray}
where $\Lambda \simeq \rho_e^{1/4}$ is the energy scale of inflation, and
\begin{equation}
T_{eq} = 3404 T_0 = 9295\ \mathrm{K} = 8.01 \times 10^{-10}\ \mathrm{GeV}.
\end{equation}

While this is a simplified expression relative to the most general case, it will be sufficient for our purpose here, since considering more general reheating models will only increase the associated uncertainty. The reheat temperature can be at most equal to the energy density, $T_{RH} \leq \Lambda$, so there is an upper bound on $N\left(k\right)$,
\begin{equation}
    \label{eq:NvsTUB}
    N\left(k\right) \leq - \ln\left(\frac{k}{a_0 H_0}\right) +  \ln\left(\frac{\Lambda}{10^{15}\ \mathrm{GeV}}\right) + 60.4.
\end{equation}
We can similarly place a lower bound on $N\left(k\right)$ by taking the reheat temperature to be at least the temperature of nucleosynthesis, $T_{RH} \gtrsim 10^{-1}\ \mathrm{GeV}$, so that
\begin{equation}
    \label{eq:NvsTLB}
    N\left(k\right) \geq - \ln\left(\frac{k}{a_0 H_0}\right) + \frac{2}{3} \ln\left(\frac{\Lambda}{10^{15}\ \mathrm{GeV}}\right) + 48.1.
\end{equation}

\begin{figure}
\includegraphics[width=3.25in]{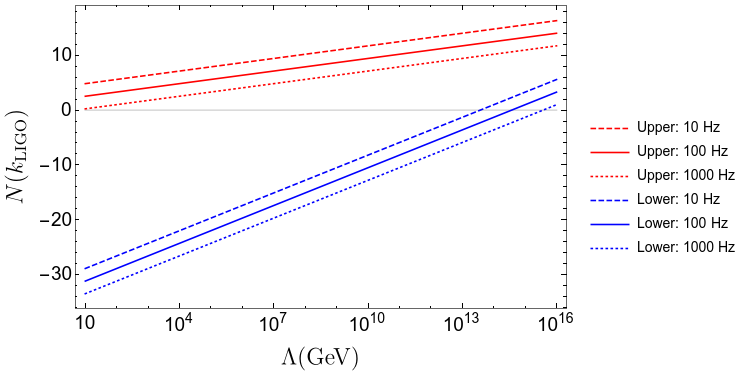}
\caption{The number of e-folds at horizon exit $N(k_{\mathrm{LIGO}})$ plotted as a function of inflationary energy scale $\Lambda$, for frequencies $\nu = 10\ \mathrm{Hz}$, $\nu = 100\ \mathrm{Hz}$, and $\nu = 1000\ \mathrm{Hz}$. Red shows the upper bound (Eq. \ref{eq:NvsTUB}), and blue shows the lower bound (Eq. \ref{eq:NvsTLB}). Regions where $N\left(k\right) < 0$ indicate no production of gravitational waves at those frequencies.}
\label{fig:LIGONvsT}
\end{figure}

\begin{figure}
\includegraphics[width=3.25in]{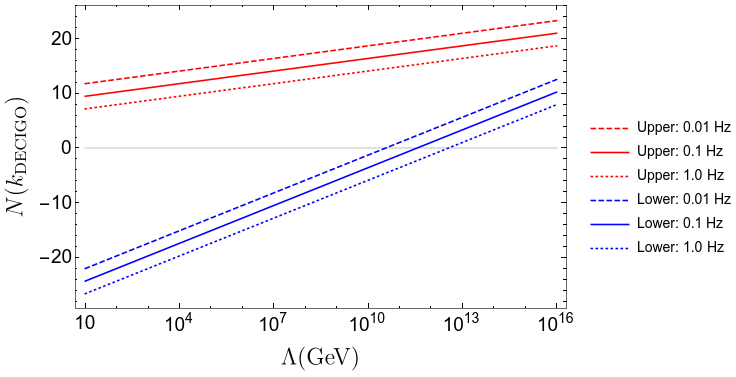}
\caption{The number of e-folds at horizon exit $N(k_{\mathrm{DECIGO}})$ plotted as a function of inflationary energy scale $\Lambda$, for frequencies $\nu = 10\ \mathrm{Hz}$, $\nu = 100\ \mathrm{Hz}$, and $\nu = 1000\ \mathrm{Hz}$. Red shows the upper bound (Eq. \ref{eq:NvsTUB}), and blue shows the lower bound (Eq. \ref{eq:NvsTLB}). Regions where $N\left(k\right) < 0$ indicate no production of gravitational waves at those frequencies.}
\label{fig:DECIGONvsT}
\end{figure}

\begin{figure}
\includegraphics[width=3.25in]{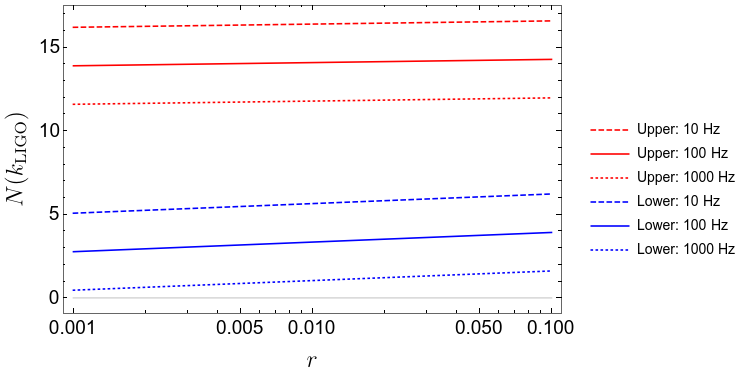}
\caption{The number of e-folds at horizon exit $N(k_{\mathrm{LIGO}})$ plotted as a function of tensor/scalar ratio $r$, for frequencies $\nu = 10\ \mathrm{Hz}$, $\nu = 100\ \mathrm{Hz}$, and $\nu = 1000\ \mathrm{Hz}$. Red shows the upper bound (Eq. \ref{eq:NvsTUB}), and blue shows the lower bound (Eq. \ref{eq:NvsTLB}). }
\label{fig:LIGONvsr}
\end{figure}

\begin{figure}
\includegraphics[width=3.25in]{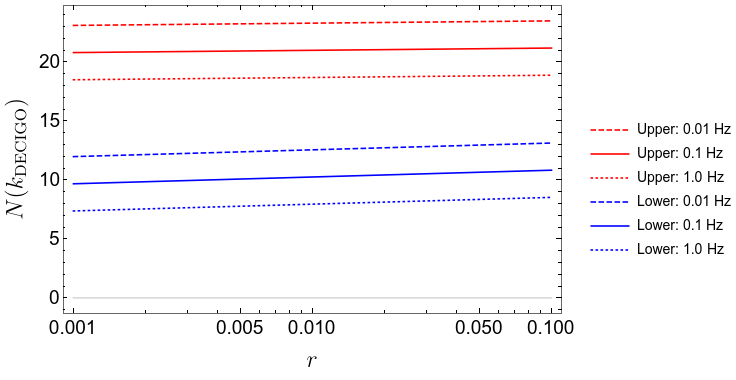}
\caption{The number of e-folds at horizon exit $N(k_{\mathrm{DECIGO}})$ plotted as a function of tensor/scalar ratio $r$, for frequencies $\nu = 0.01\ \mathrm{Hz}$, $\nu = 0.1\ \mathrm{Hz}$, and $\nu = 1.0\ \mathrm{Hz}$. Red shows the upper bound (Eq. \ref{eq:NvsTUB}), and blue shows the lower bound (Eq. \ref{eq:NvsTLB}). }
\label{fig:DECIGONvsr}
\end{figure}

Figure \ref{fig:LIGONvsT} shows upper and lower bounds on $N\left(k_{\mathrm{LIGO}}\right)$ versus inflationary energy scale $\Lambda$, and Fig. \ref{fig:DECIGONvsT} shows $N\left(k_{\mathrm{DECIGO}}\right)$. The first thing to note is that for low scales of the inflationary energy density, $N\left(k\right)$ actually becomes \textit{negative}, which means that those modes never leave the horizon during inflation, and there is no production of gravitational waves at all! However, the energy scale of inflation and the tensor/scalar ratio at CMB scales are directly related, 
\begin{equation}
\Lambda \simeq r^{1/4} \times \left(3.3 \times 10^{16}\ \mathrm{Gev}\right),
\end{equation}
so that the \textit{ansatz} of $r_{0.002} > 0.001$, implies $\Lambda \geq 5.9 \times 10^{15}\ \mathrm{GeV}$. Figure \ref{fig:LIGONvsr} shows bounds on $N\left(k_{\mathrm{LIGO}}\right)$ for the range $r = [0.001,0.1]$. In this case we see that primordial perturbations corresponding to LIGO sensitivity exit the horizon during inflation within a few e-folds of the end of inflation and the onset of reheating. This is precisely the region where the assumption of slow roll breaks down, and we expect parameters such as the expansion rate $H$ and the slow roll parameter $\epsilon$ to be varying relatively rapidly. In this case, we no longer expect power-law behavior in the primordial tensor spectrum, and there will be a corresponding systematic uncertainty in the amplitude of tensors at direct detection frequencies. Figure \ref{fig:DECIGONvsr} shows $N\left(k_{\mathrm{DECIGO}}\right)$ for the planned DECIGO peak sensitivity range $\nu = 0.01 - 1.0\ \mathrm{Hz}$ \cite{Kawamura:2020pcg}, or $k_{\mathrm{DECIGO}} = 6.47 \times 10^{13}\ \mathrm{Mpc}^{-1}$ for $\nu = 0.1\ \mathrm{Hz}$.

In the next section, I use the inflationary flow formalism to show that extrapolating an exact power-law spectrum to direct detection scales from CMB scales results in a systematic \textit{overestimate} of the tensor amplitude at direct detection scales by as much as two orders of magnitude, relative to realistic single-field inflation models. 

\section{Flow Analysis}

The inflationary flow formalism \cite{Hoffman:2000ue,Kinney:2002qn} is a convenient method of generating large ensembles of single-field inflation models consistent with a particular set of observational constraints \cite{Easther:2002rw}. The equations of motion for the spacetime and scalar field are given by:
\begin{eqnarray}
\label{eq:Friedmann}
&&H^2 = \left(\frac{\dot a}{a}\right)^2 = \frac{1}{3 M_{\mathrm{P}}^2}\left[\frac{1}{2} \dot\phi^2 + V\left(\phi\right)\right],\cr
&&\ddot \phi + 3 H \dot\phi + V'\left(\phi\right) = 0. 
\end{eqnarray}
If the field evolution is monotonic in time, we can write the scale factor $a\left(\phi\right)$ and Hubble parameter $H\left(\phi\right)$ as functions of the field $\phi$ rather than time. Equations (\ref{eq:Friedmann})  can then be re-written exactly in the Hamilton-Jacobi form
\begin{eqnarray}
\label{eq:Hamjacobi}
V\left(\phi\right) &=& 3 M_{\mathrm{P}}^2 H^2\left(\phi\right) \left[1 - \frac{2 M_{\mathrm{P}}^2}{3} \left(\frac{H'\left(\phi\right)}{H\left(\phi\right)}\right)^2\right],\cr
\dot\phi &=& - 2 M_{\mathrm{P}}^2 H'\left(\phi\right).
\end{eqnarray} 
The first slow roll parameter $\epsilon$ is related to the equation of state by
\begin{equation}
\epsilon = \frac{3}{2} \left(1 + w\right) = -\frac{a}{H} \frac{d H}{d a}.
\end{equation}
Inflation then occurs for $w < -1/3$, or $\epsilon < 1$. During inflation, the scale factor increases quasi-exponentially, with the Hubble parameter $H \simeq {\rm const.}$ and
\begin{equation}
a \propto \exp\left[\int{H dt}\right] \equiv e^{-N}.
\end{equation}
Here $N$ is the number of e-folds before the end of inflation, which is related to the field $\phi$ by
\begin{eqnarray}
\label{eq:numberofefolds}
d N \equiv - H dt &=& -\frac{d a(\phi)}{a(\phi)} \cr
 &=& \frac{1}{\sqrt{2} M_{\mathrm{P}}}\frac{d\phi}{\sqrt{\epsilon(\phi)}}.
\end{eqnarray}
We can then write the parameter $\epsilon$ as
\begin{equation}
\epsilon = \frac{1}{H} \frac{d H}{d N} = 2 M_{\mathrm{P}}^2 \left(\frac{H'\left(\phi\right)}{H\left(\phi\right)}\right)^2.
\end{equation}
We define an infinite hierarchy of \textit{Hubble slow roll parameters} \cite{Copeland:1993jj,Liddle:1994dx} by taking successive derivatives of the Hubble parameter $H$ with respect to the field $\phi$:
\begin{eqnarray}
\label{eq:definflationparamshierarchy}
\epsilon &\equiv& {2 M_{P}^2 } \left(\frac{H'(\phi)}{H(\phi)}\right)^2, \cr 
\eta &\equiv& {2 M_{P}^2 } \frac{H''(\phi)}{H(\phi)}, \cr
{}^{2} \lambda_H  &\equiv& {4 M_{P}^4}  \frac{H'(\phi) H'''(\phi)}{H^2(\phi)}, \cr 
\vdots \cr   
{}^\ell \lambda_H &\equiv& {\left(2 M_{P}^2\right) ^ \ell} \frac{H'(\phi)^{\left(\ell-1\right)}}{H(\phi)^ \ell} \frac{d^{\left(\ell +1\right)} H(\phi)}{d \phi ^ {\left(\ell +1\right)}},
\end{eqnarray}
Using Eq. (\ref{eq:numberofefolds}) to take derivatives of $\epsilon$ with respect to $N$, we can generate an infinite set of differential equations relating the parameters:
\begin{eqnarray} 
\label{eq:flowequations}
\frac{d \epsilon}{d N} &=& 2 \epsilon \left(\eta - \epsilon\right), \cr
\frac{d \eta}{d N} &=&  - \epsilon \eta + {}^{2} \lambda_H , \cr
\vdots \cr  
\frac{d {}^\ell \lambda_H} {d N} &=& \left[(\ell-1) \eta - \ell \epsilon \right] {}^ \ell \lambda_H + {} ^ {(\ell + 1)} \lambda_H,
\end{eqnarray}
This hierarchy of flow equations completely specifies the evolution of the spacetime \cite{Kinney:2002qn}: once the parameters $\epsilon\left(N\right)$, $\eta\left(N\right)$ and so forth are known, the cosmological dynamics and scalar field potential are fixed via the Hamilton-Jacobi Equations (\ref{eq:Hamjacobi}). In practice, the hierarchy of equations must be truncated at finite order, which I take to be $\ell \leq 6$. Since the system of equations is first-order, specifying a point in the parameter space is sufficient to fully fix the solution. (Note that solutions to the truncated system of equations are also exact solutions to the \textit{infinite} system.)

To test the validity of a power-law extrapolation from CMB to direct detection scales, I perform a Monte Carlo evaluation of the flow equations (\ref{eq:flowequations}), collecting 500 realizations consistent with Planck+BICEP/Keck 95\% confidence bounds on the scalar spectral index $n_S = [0.9569,0.9733]$, $r_{k=0.002} < 0.056$ , with the scalar power spectrum normalized at the pivot scale to $A_S = 2.090524 \times 10^{-9}$ \cite{Akrami:2018odb}. I take $N(k = 0.002)$ varying between 50 and 60, to ensure that the modes at short wavelength are redshifted fully into the infrared limit, and make the additional \textit{ansatz} of selecting only realizations for which $r_{0.002} \geq 0.001$, consistent with the expected sensitivity of near-future CMB polarization measurements such as CMB-S4 \cite{2019BAAS...51g.209C}. 

\begin{figure}
\includegraphics[width=3.25in]{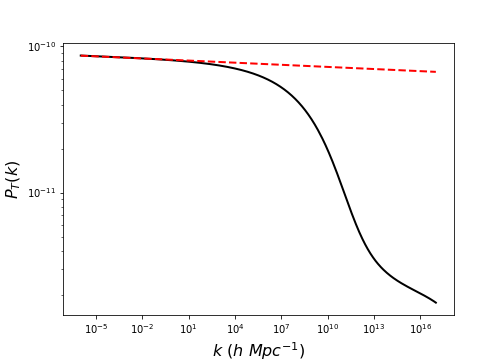}
\caption{The exact tensor power spectrum (solid black line) vs. the power-law extrapolation (dotted red line) for an example inflationary potential. The power-law form associated with slow roll is a good approximation at small $k$, but the exact form deviates strongly from power-law near the end of inflation, at large $k$.}
\label{fig:PT}
\end{figure}

\begin{figure}
\includegraphics[width=3.25in]{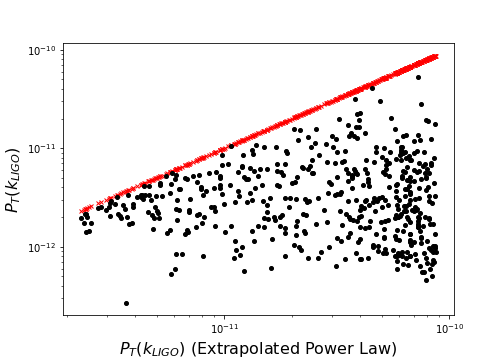}
\caption{The primordial tensor amplitude $P_T\left(k_{\mathrm{LIGO}}\right)$ for an exact slow roll solution vs. a power-law extrapolation from the CMB pivot scale $k = 0.002\ \mathrm{Mpc}^{-1}$ to the LIGO/Virgo scale $k_{\mathrm{LIGO}} = 6.47 \times 10^{16}\ \mathrm{Mpc}^{-1}$. The crosses show the extrapolated power-law, which forms a line of unit slope. The points show the calculated values for the exact slow roll potentials, which lie as much as two orders of magnitude below.}
\label{fig:LIGOall}
\end{figure}

\begin{figure}
\includegraphics[width=3.25in]{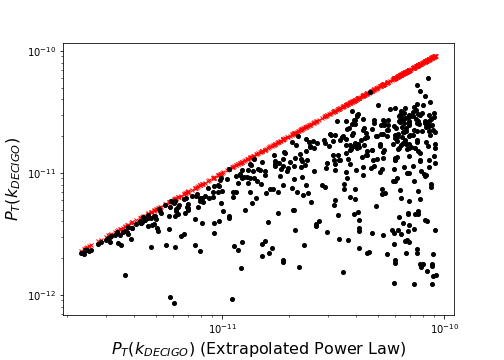}
\caption{The primordial tensor amplitude $P_T\left(k_{\mathrm{DECIGO}}\right)$ for an exact slow roll solution vs. a power-law extrapolation from the CMB pivot scale $k = 0.002\ \mathrm{Mpc}^{-1}$ to the DECIGO scale $k_{\mathrm{DECIGO}} = 6.47 \times 10^{13}\ \mathrm{Mpc}^{-1}$. The crosses show the extrapolated power-law, which forms a line of unit slope. The points show the calculated values for the exact slow roll potentials, which lie as much as two orders of magnitude below.}
\label{fig:DECIGOall}
\end{figure}

Comparison of primordial tensor modes to direct detection bounds on the energy density in gravitational waves requires application of a linear transfer function $T(k)$ to the primordial spectrum, taking into account cosmological evolution after wave modes becomes subhorizon, which is done for an ensemble of inflationary potentials, for example, in Ref. \cite{Caligiuri:2014ola}. Since our goal here is to compare a power-law extrapolation to realistic single-field inflation models, this step is not necessary: it is sufficient to compare the amplitudes of the \textit{primordial} spectra directly at direct detection scales, since linearity guarantees that any systematic error will be retained after processing by the transfer function. Both the background evolution and the tensor power spectrum are calculated numerically, without using the slow roll approximation. The result is shown in Fig. \ref{fig:LIGOall} for LIGO/Virgo, and Fig. \ref{fig:DECIGOall} for DECIGO. Realistic inflation models lie as much as two orders of magnitude below the expectation from extrapolation of a power-law from CMB scales, because the Hubble parameter typically decreases rapidly near the end of inflation. This is shown for an example case in Fig. \ref{fig:PT}. The exact single-field solution differs strongly from the power-law extrapolation at large $k$.

\begin{figure}
\includegraphics[width=3.25in]{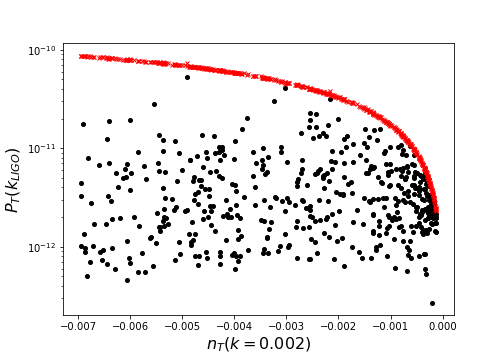}
\caption{The primordial tensor amplitude $P_T\left(k_{\mathrm{LIGO}}\right)$ at LIGO/Virgo scales plotted vs. the tensor spectral index at the CMB pivot scale $k = 0.002\ \mathrm{Mpc}^{-1}$. The red crosses show the values for a power-law extrapolation of the tensor power spectrum, which are highly correlated due to the single-field consistency condition. The black points show the values for the exact slow roll calculation, which are uncorrelated with $n_T$ measured at the CMB pivot scale.}
\label{fig:LIGOnT}
\end{figure}

\begin{figure}
\includegraphics[width=3.25in]{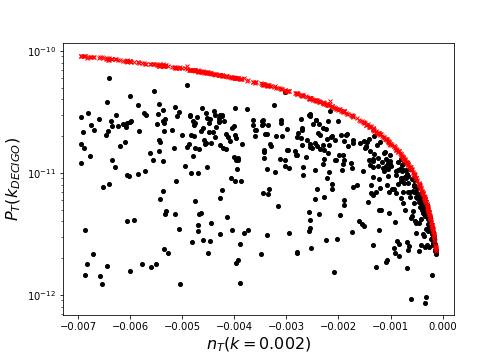}
\caption{The primordial tensor amplitude $P_T\left(k_{\mathrm{DECIGO}}\right)$ at DECIGO scales plotted vs. the tensor spectral index at the CMB pivot scale $k = 0.002\ \mathrm{Mpc}^{-1}$. The red crosses show the values for a power-law extrapolation of the tensor power spectrum, which are highly correlated due to the single-field consistency condition. The black points show the values for the exact slow roll calculation, which are uncorrelated with $n_T$ measured at the CMB pivot scale.}
\label{fig:DECIGOnT}
\end{figure}

Figure \ref{fig:LIGOnT} shows the primordial tensor amplitude $P_T\left(k_{\mathrm{LIGO}}\right)$ plotted versus the tensor spectral index at the CMB pivot scale, $n_T\left(k_{0.002}\right)$. The corresponding result for DECIGO is plotted in Fig. \ref{fig:DECIGOnT}. Because the tensor/scalar ratio $r$ and the spectral index $n_T$ are related by the consistency relation (\ref{eq:consistency}), we expect the tensor amplitude $P_T$ and the spectral index $n_T$ to be highly correlated, which is in fact the case for the power-law extrapolation. This correlation, however, completely disappears for realistic inflationary potentials. This means that, for all practical purposes, the tensor amplitude at direct detection scales gives no information about the tensor spectral index at CMB scales. 

\section{Conclusions}

Primordial tensor perturbations at superhorizon scales are a \textit{unique} prediction of cosmological inflation \cite{Geshnizjani:2014bya}. The single-field consistency condition (\ref{eq:consistency}) relating the tensor amplitude to the slope of the power spectrum is a key target for observational tests of inflation. CMB bounds combined with high-frequency gravitational wave constraints from direct detection bounds such as the LIGO/Virgo and DECIGO observatories in principle provide a large lever arm with which to constrain the slope of the primordial spectrum. Current bounds are complicated by the fact that no primordial gravitational waves have yet been detected, and it is known to be somewhat difficult to measure the slope of a function without having measured its value at any point. In fact, for low enough inflationary energy scale, no primordial tensor modes at all are produced at direct detection scales. Any constraint must therefore be model-dependent, for example by assuming a lower bound on the tensor amplitude at CMB scales by \textit{ansatz}. Inflation predicts an approximately power-law spectrum of primordial perturbations over at least a few orders of magnitude in scale, and it is typically assumed that this power-law behavior extends to frequencies accessible by direct detection experiments. 

In this paper, I examine the consistency of power-law extrapolation of the inflationary tensor spectrum from CMB scales to direct detection scales, and find that such extrapolation overestimates the direct detection signal by up to two orders of magnitude relative to an ensemble of realistic single-field inflation models with $r_{0.002} > 0.001$ (Figs. \ref{fig:LIGOall}, \ref{fig:DECIGOall}). This is because high-frequency modes are produced very near the end of inflation, when the Hubble parameter is rapidly decreasing, and the tensor amplitude is correspondingly suppressed (Fig. \ref{fig:PT}). In addition, there is no correlation between $n_T$ at CMB scales and the tensor amplitude at LIGO/Virgo scales (Fig. \ref{fig:LIGOnT}), and only weak correlation at DECIGO scales (Fig. \ref{fig:DECIGOnT}). This means that any constraint on $n_T$ at CMB scales, and therefore the consistency relation, is highly model-dependent. Conversely, a detection of primordial tensors at \textit{both} CMB and (for example) DECIGO scales would place a strong constraint the form of the inflationary potential $V\left(\phi\right)$ \cite{Caligiuri:2014ola}. Until then, the only well-supported conclusion we can make about the spectral index of the primordial tensor power spectrum is: we have no idea. 

\section*{Acknowledgements}

I thank Alessandro Melchiorri, Sunny Vagnozzi, and Eleonora Di Valentino for helpful discussions. The inflationary flow code was written by Brian A. Powell and WHK. 

\begin{acknowledgments}

\end{acknowledgments}

\bibliography{Paper.bib}

\end{document}